\documentclass[pra,twocolumn,superscriptaddress,footinbib,bibliography]{revtex4-2}
\usepackage{graphicx}% Include figure files
\usepackage{dcolumn}% Align table columns on decimal point
\usepackage{bm}% bold math
\usepackage{amsmath}
\usepackage{amssymb}
\usepackage{latexsym}
\usepackage{amsthm}
\usepackage{array}
\usepackage{amsfonts}
\usepackage{mathrsfs}
\usepackage{xcolor}
\usepackage{kotex}
\usepackage{verbatim}
\usepackage[caption=false]{subfig}
\usepackage{natbib}
\usepackage{enumitem}
\usepackage{bbold}
\usepackage{ulem}
\usepackage{cancel}
\usepackage[colorlinks=true,urlcolor=blue,citecolor=blue,linkcolor=blue]{hyperref}

\begin{document}
%\fontfamily{phv}\selectfont

\title{Contextual quantum metrology}

\author{Jeongwoo Jae}
\affiliation{Department of Physics, Hanyang University, Seoul, 04763, Republic of Korea}

\author{Jiwon Lee}
\affiliation{Department of Physics, Hanyang University, Seoul, 04763, Republic of Korea}

\author{M. S. Kim}
\affiliation{QOLS, Blackett Laboratory, Imperial College London, London SW7 2AZ, United Kingdom}

\author{Kwang-Geol Lee}
\email{kglee@hanyang.ac.kr}
\affiliation{Department of Physics, Hanyang University, Seoul, 04763, Republic of Korea}

\author{Jinhyoung Lee}
\email{hyoung@hanyang.ac.kr}
\affiliation{Department of Physics, Hanyang University, Seoul, 04763, Republic of Korea}

\begin{abstract}
Quantum metrology promises higher precision measurements than classical methods. Entanglement has been identified as one of quantum resources to enhance metrological precision. However, generating entangled states with high fidelity presents considerable challenges, and thus attaining metrological enhancement through entanglement is generally difficult. Here, we show that contextuality of measurement selection can enhance metrological precision, and this enhancement is attainable with a simple linear optical experiment. We call our methodology ``contextual quantum metrology" (coQM). Contextuality is a nonclassical property known as a resource for various quantum information processing tasks. Until now, it has remained an open question whether contextuality can be a resource for quantum metrology. We answer this question in the affirmative by showing that the coQM can elevate precision of an optical polarimetry by a factor of $1.4$ to $6.0$, much higher than the one by quantum Fisher information, known as the limit of conventional quantum metrology. We achieve the contextuality-enabled enhancement with two polarization measurements which are mutually complementary, whereas, in the conventional method, some optimal measurements to achieve the precision limit are either theoretically difficult to find or experimentally infeasible. These results highlight that the contextuality of measurement selection is applicable in practice for quantum metrology.
\end{abstract}

\maketitle

\newcommand{\bra}[1]{\langle {#1}\vert} 
\newcommand{\ket}[1]{\vert {#1}\rangle} 
\newcommand{\abs}[1]{\vert {#1} \vert} 
\newcommand{\avg}[1]{\langle {#1}\rangle}
\newcommand{\braket}[2]{\langle {#1} \vert {#2} \rangle}
\newcommand{\commute}[2]{\left[{#1},{#2}\right]}
\newcommand{\mdo}[1]{\left[#1\right]}
\newcommand{\rmk}[1]{\textcolor{violet}{#1}}
\newcommand{\del}[1]{\textcolor{violet}{\sout{#1}}}
\newcommand{\add}[1]{\textcolor{blue}{\uwave{#1}}}
\newcommand{\sep}[1]{\left[ {#1}\right]} 
\newcommand{\notimplies}{%
  \mathrel{{\ooalign{\hidewidth$\not\phantom{=}$\hidewidth\cr$\implies$}}}}
\newtheorem{theorem}{Theorem}
\newcommand{\E}[1]{\mathbf{E}[#1]}
\newcommand{\Var}[1]{\mathbf{Var}[#1]}
%---------------------------------------------------------------------------------------
%---------------------------------------------------------------------------------------

\section*{Introduction}
Precision measurement has played a crucial role in the development of natural science and engineering since measurement is a means for observing nature. As a technology for precision measurement, quantum metrology has recently drawn attention with a wide range of applications such as microscopy~\cite{Casacio2021}, imaging~\cite{Treps2002,Brida2010}, patterning~\cite{Boto2000,Parniak2018}, gravitational wave detection~\cite{Abramovici1992,LIGO2011,Aasi2013}, and time keeping~\cite{Giovannetti2001,Pedrozo2020}. Quantum metrology enables measurements going beyond precision of the standard quantum limit which can be obtained from the most-classical state in quantum physics. One of the resources for the precision enhancement is entanglement, a nonclassical property of quantum states~\cite{Giovannetti2004,Giovannetti2006,Giovannetti2011,Tan2019}. However, an entangled state can easily lose its property through interaction with other objects, while the interaction is inevitable in metrology. This makes it challenging to generate and manipulate an entangled state. Due to the limitations, it is difficult in practice to attain the entanglement-enabled enhancement of precision. If easy-to-implement resources for metrology are found, the performance of quantum metrology can be greatly enhanced, as well as its practicality. In this work, we demonstrate that contextuality of measurement selection~\cite{Ryu_2019}, another nonclassical property, is an easy-to-implement resource for quantum metrology.

Specifically, contextuality here refers to dependency of quantum systems on measurement context~\cite{Spekkens2005}. Unlike classical predictions, quantum predictions for a given measurement can change depending on whether another measurement is performed simultaneously or not. Bell-Kochen-Specker theorem first showed that quantum physics is contextual~\cite{Bell64,Kochen1967}, and this has been experimentally proved on various quantum systems~\cite{Yuji2006,Kirchmair2009,Jerger2016,Aonan2019}. Also, it has been revealed that the contextuality can be a resource for quantum information processing tasks such as quantum key distribution~\cite{Antonio2007,Reichardt2013}, universal quantum computing~\cite{Howard2014}, quantum state discrimination~\cite{Schmid2018}, and quantum machine learning~\cite{Gao2022,Anschuetz2023}. Yet, whether the contextuality can be a resource for quantum metrology remains a question.

To demonstrate the precision enhancement from the contextuality of measurement selection, we propose a method which we call contextual quantum metrology (coQM). Unlike conventional quantum metrology, the coQM utilizes two measurement settings and their contextuality. In our experiment, we adopt an optical polarimetry devised to measure concentration of sucrose solution~\cite{Yoon2020}, and modify its scheme for the coQM. Our experiment employs two polarization measurements in mutually unbiased (or complementary) bases, and their selection context is implemented by toggling a polarizing beam splitter `in' and `out' from its optical path. Our setup is scalable in that the size of the experiment does not increase along with the increase of the number of probe photons. Also, the enhanced precision is attainable without error correction or mitigation which requires overhead~\cite{Zhou2018,Maciejewski2020}. We finally show that the precision of coQM can go beyond the precision limit of conventional quantum metrology~\cite{Helstrom1969,Holevo2011,Braunstein1994} by a factor of $1.4$ to $6.0$.

\begin{figure*}[t!]
    \centering
    \includegraphics[width=0.9\textwidth]{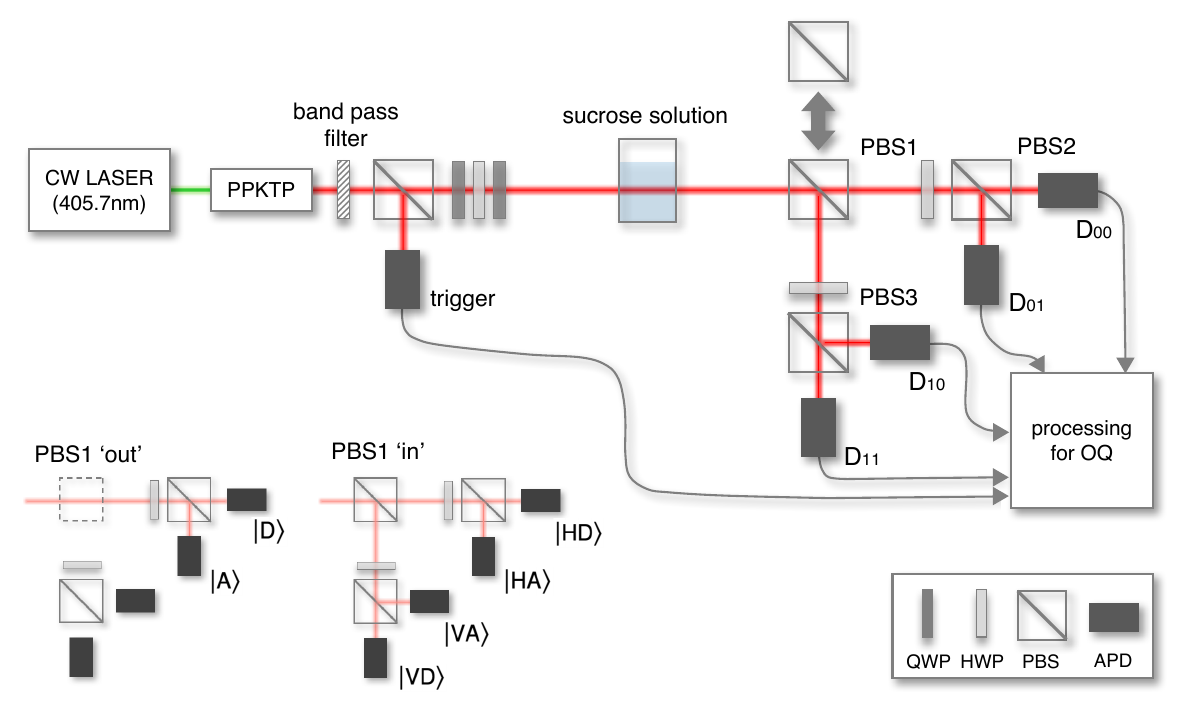}
    \caption{{\bf An experimental schematic for coQM.} Our probe state is a polarization state of a heralded single-photon source (see Methods). The probe polarization rotates by angle $\alpha l c$ when it propagates through the sucrose solution, where $\alpha \approx 34.1$ deg ml dm$^{-1}$ g$^{-1}$ is the specific rotation of the sucrose solution, $l=0.1$ dm is traveling length of light in the solution, and $c$ g ml$^{-1}$ is the  concentration of the solution. We estimate the concentration $c$ by measuring the polarization change. Here, we consider two measurement settings $A$ and $B$, where $A$ measures polarization in $H/V$ basis and $B$ does in a $D/A$ basis tilted $45^\circ$ from the basis of $A$ by half-wave plate (HWP). For triggered event, the probabilities of the polarization bases are determined from relative counts of four avalanche photodiodes (APDs), $D_{00}$, $D_{01}$, $D_{10}$, and $D_{11}$. When polarizing beam splitter PBS$1$ is ‘out’, the measurement setup corresponds to the $B$ measurement, and the counts on $D_{00}$ and $D_{01}$ determine probabilities of $\ket{D}$ and $\ket{A}$, respectively. When the PBS$1$ is ‘in’, the measurement setup corresponds to the consecutive measurement performing $A$ first and $B$ later, and the counts of $D_{00}$, $D_{01}$, $D_{10}$, and $D_{11}$ represent joint probabilities of $\ket{HD}$, $\ket{HA}$, $\ket{VA}$, and $\ket{VD}$, respectively. We combine measuring data to construct operational quasiprobabiltiy (OQ) in Eq.~\eqref{eq:OQ_theory}, and an estimate is calculated with maximum likelihood estimator of OQ in Eq.~\eqref{eq:mlew}. The coQM utilizes the context of selecting the measurement $A$ to enhance its precision. Experimental results in Fig.~\ref{fig:result} demonstrate the enhancement.}
    \label{fig:experiment}
\end{figure*}

\section*{Contextual quantum metrology}
Fig.~\ref{fig:experiment} shows an experimental schematic for the coQM. Here, the coQM estimates the concentration of sucrose solution by following four steps: preparing a polarized single photon as a probe light (see Methods), interacting the photon with the sucrose solution, measuring the polarization, and calculating an estimate via maximum likelihood estimator (MLE) using the operational quasiprobability which will be discussed later as in Eq.~\eqref{eq:OQ_theory}. The photon interacts with the sucrose solution as it propagates through the solution. Afterwards, the photon polarization rotates by angle $\alpha c l$, where $\alpha$, $c$, and $l$ are the specific rotation of the sucrose solution, the concentration of the solution, and traveling length of light in the solution, respectively~\cite{Yoon2020}. We can decide the concentration by measuring the polarization change for a given specific rotation and traveling length. All the procedures here look similar to those of the conventional quantum metrology~\cite{Liu_2020}, but the measurement and the estimation are steps that differentiate the coQM from the conventional approach.

In the measurement step, the coQM employs two different polarization measurement settings $A$ and $B$ to utilize the contextuality of measurement selection therefrom~\cite{Ryu_2019}. $A$ measures the polarization in a specific basis $\{\ket{H},\ket{V}\}$ and $B$ does in a basis $\{\ket{D},\ket{A}\}$ tilted $45^{\circ}$ from the basis of $A$ by half-wave plate. Our basic setup is the measurement by $B$, and we consider a context of whether measurement $A$ is selected to be performed or not, prior to performing $B$. If $A$ is not selected, the measurement $B$ only is performed, and probability is given by $p(b|B)$ for a binary value $b$. If $A$ is selected, the experimental setup runs the consecutive measurement performing $A$ first and $B$ later. In this case, probability is given by $p(a,b|AB)$ for an outcome pair $(a,b)$, where $a$ is a binary outcome of $A$. In our experiment (Fig.~\ref{fig:experiment}), the context of selecting $A$ is implemented by toggling `in' and 'out' the state of polarizing beam splitter PBS$1$. Depending on the context, the quantum prediction of measurement $B$ changes, so the two probabilities $p(b|B)$ by $B$ only and $p(b|AB)=\sum_a p(a,b|AB)$ by the consecutive differ in general, i.e., $p(b|B)\neq p(b|AB)$~\cite{Leggett1985}. (We will discuss that this difference stems from the incompatibility of measurements~\cite{Busch86}.) Our metrology utilizes this effect to enhance precision of the polarimetry.

In the estimation step, we employ a noncontextual (or context-free) model, so-called operational quasiprobability~\cite{Ryu2013,Jae2017}, which is given by
\begin{eqnarray}
\label{eq:OQ_theory}
w(a,b) &=& p(a,b|AB) + \frac{1}{2}\left( p(b|B) - p(b|AB) \right).
\end{eqnarray}
The context-free condition in our measurement setup is to assume that the prediction of measurement $B$ is invariant under the context of selecting the measurement $A$. This is called the condition of no-signaling in time represented by $p(b|B) = p(b|AB)$, $\forall b$~\cite{Leggett1985}. The crucial property of the operational quasiprobability is that, for the context-free condition, $w$ is reduced to the probability by the consecutive measurement, $w(a, b) = p(a,b|AB)$, $\forall a,b$. To the contrary, the quantum predictions violate the condition in general, $w(a,b) \ne p(a,b|AB)$, and $w(a,b)$ can even be negative-valued~\cite{Ryu2013,Jae2017}.

The conventional quantum metrology estimates a physical parameter $\theta$ with an estimator based on a conditional probability $p(x|\theta)$ for a data set $\{x_i\}_{i=1}^{N_s}$ (see Supplementary Information). In the coQM, the operational quasiprobability plays a role of the conditional probability for the two data sets ${\bf x}_B = \{b_i\}_{i=1}^{N_s}$ and ${\bf x}_{AB}=\{(a_j,b_j)\}_{j=1}^{N_s}$. In other words, the coQM calculates an estimate of polarization $\check{\theta}$ with a maximum likelihood estimator given by
\begin{equation}
\label{eq:mlew}
 \check{\theta} ~~\text{s.t.}~~ \partial_\theta l_W(\theta|{\bf x}_B,{\bf x}_{AB}) = 0,
\end{equation}
where $l_W(\theta|{\bf x}_B,{\bf x}_{AB})$ is a log-likelihood function for $w$ (see Methods). The possible problem caused by this replacement is that $w$ can be negative unlike the conditional probability, so that the log-likelihood function diverges. However, we find that $w$ remains positive for some range of parameters. We focus on the case for $w$ to be applicable to the log-likelihood function without the divergence problem. Finally, for an initial polarization $\theta_0$, we derive the estimate of concentration $\check{c}$ from the polarization change as $\check{c}=(\check{\theta}-\theta_0)/\alpha l$.

\section*{Results}
Our goal is to demonstrate outperformance of the coQM over the conventional quantum metrology. We employ error of estimate $\Delta \theta$, the standard deviation that the estimate differs from the actual value, to quantify the performance of estimation. The smaller the error is, the more precise the estimate is.

As a baseline of performance, we take the conventional quantum metrology and its error given by quantum Fisher information (QFI) $F_q$, $\Delta \theta_q={1}/{\sqrt{N_s F_q}}$, where $N_s$ is the number of samples. This is known as the lower bound of error in the conventional method. In the coQM, we propose contextual Fisher information (coFI) to quantify the performance of the coQM,
\begin{eqnarray}
F_\mathrm{co} := \sum_{ab} w(a,b|\theta) \left( \frac{\partial \log w(a,b|\theta)}{\partial \theta} \right)^2.
\end{eqnarray}
In the asymptotic limit of $N_s \rightarrow \infty$, the error of the coQM $\Delta \theta_\mathrm{co}$ approaches to $1/\sqrt{N_s F_\mathrm{co}}$ (see Supplementary Information for the asymptotic property and estimator of coFI).

{\bf Contextuality-enabled enhancement} The coQM gains precision enhancement over the conventional quantum metrology if
\begin{eqnarray}
\label{eq:enhancement}
\Delta \theta_\mathrm{co} < \frac{\Delta\theta_q}{\sqrt{2}}.
\end{eqnarray}
Our method uses the two data sets ${\bf x}_B$ and ${\bf x}_{AB}$. If each data set collects $N_s$ samples, the total number of samples is $2N_s$ in our method. Reduction factor $\sqrt{2}$ in the error by the conventional is introduced, assuming the conventional takes $2N_s$ samples (which is equivalent to comparing $F_\mathrm{co}$ to $2 F_q$).

\begin{figure*}[t!]
    \centering
    \includegraphics[width=1\textwidth]{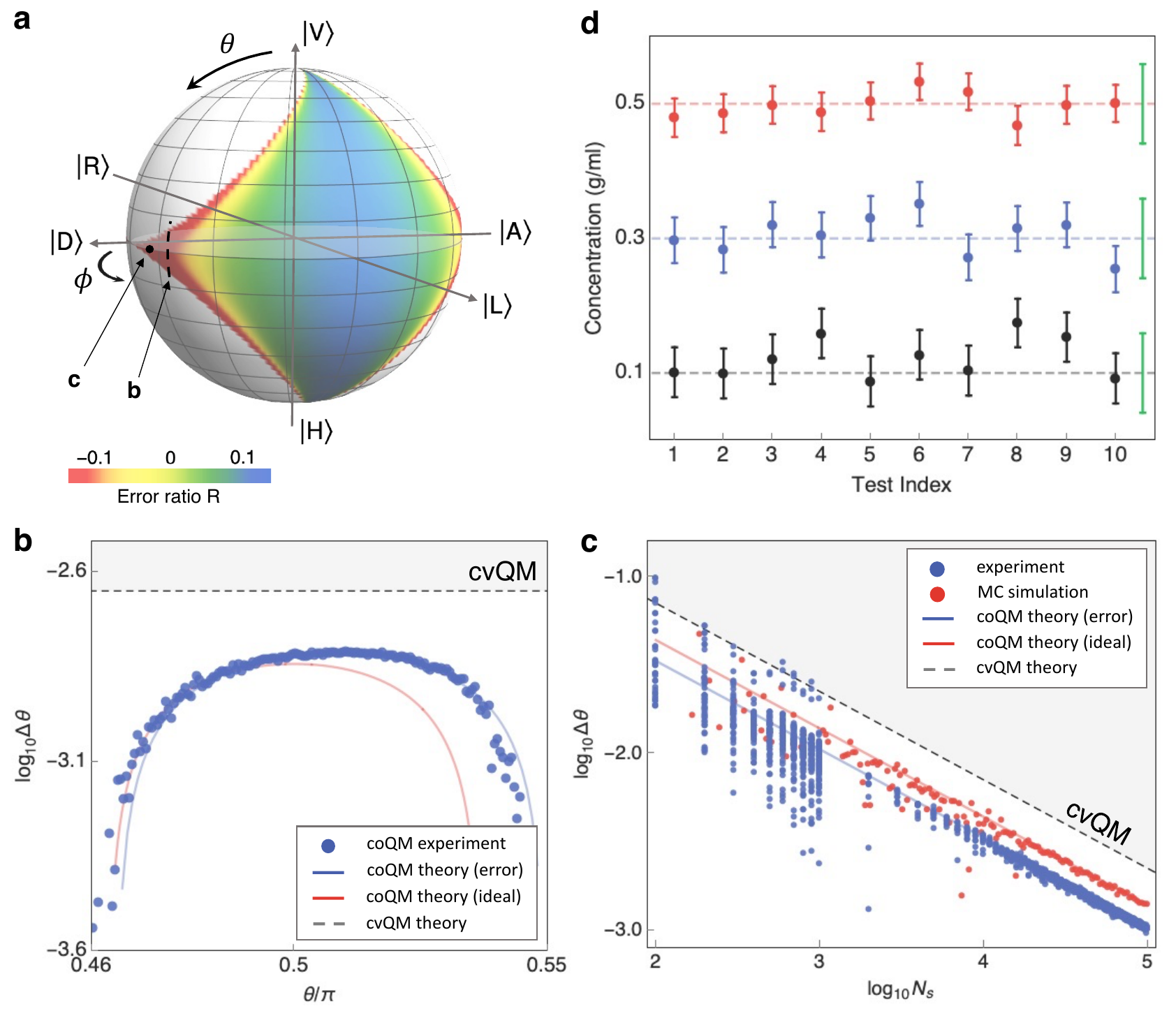}
    \caption{{\bf Simulation and experiment results of the coQM.} {\bf a} Landscape of error ratio of our $\Delta \theta_\mathrm{co}$ to the conventional $\Delta\theta_q/\sqrt{2}$, $R:=\log_{10}\left(\sqrt{2}\Delta\theta_\text{co}/\Delta\theta_q\right)$, on the Bloch sphere of a probe state $\ket{\psi} = \cos(\theta/2) \ket{H} + e^{i\phi}\sin(\theta/2)\ket{V}$, obtained by Monte-Carlo (MC) simulation. The coQM gains the contextuality-enabled enhancement if $R<0$. The operational quasiprobability (OQ) is negative in the white regions. {\bf b} Experiment results of $\theta$ estimation. We prepare the probe states, $\ket{\psi}$s, by selecting $149$ equiangular points from range $0.46\pi\le\theta\le0.55\pi$ for a fixed $\phi=0.15\pi$. For each probe state $\ket{\psi}$, we conduct $\theta$ estimation by drawing $10^5$ samples ($N_s=10^5$). Experiments (blue circles) assert that the coQM errors $\Delta\theta_\mathrm{co}$ are smaller than the minimum error in the conventional quantum metrology (cvQM) $\Delta\theta_q/\sqrt{2} = 1/\sqrt{2N_s F_q}$ (dashed line), where quantum Fisher information $F_q=1$. The red solid line is the theoretical prediction, assuming the ideal optical setup. We characterize the actual experimental setup by applying a systematic error model and use the resultant model parameters to make the theoretical prediction (blue solid line) (see Supplementary Information). {\bf c} Sample-size dependency of estimation error $\Delta \theta$. In experiment, we prepare a probe state with $\theta=0.5\pi$ and $\phi=0.1\pi$. For small sample sizes, OQ can be negative by statistical fluctuation, which occasionally leads to estimation failures. For $N_s=10^2$, failure rate is $88\%$, whereas the failure rate becomes significantly low for larger sample sizes and it is negligible for $N_s\ge7\times10^3$. {\bf d} We estimate concentrations $c =0.1$, $0.3$, and $0.5$ g ml$^{-1}$ with a probe state $\ket{\psi}$ prepared with $\theta_0=0.5\pi$ and $\phi=0.15\pi$. We draw $N_s=10^5$ samples to estimate each concentration and repeat each estimation $10$ times. The error of concentration $\Delta c$ comes from that of polarization $\Delta\theta$ by $\Delta c = \Delta\theta/\alpha l$. Error bar at each individual point represents $\Delta c_\mathrm{co}$. The green error bar in each concentration represents the minimum error by the cvQM, $\Delta c_q/\sqrt{2}$.}
    \label{fig:result}
\end{figure*}

We here suggest specific cases satisfying the criterion~\eqref{eq:enhancement}. Instead of a theoretical proof, we briefly summarize the theory behind the enhancement of precision by following arguments: For the noncontextual model, the operational quasiprobability $w(a,b|\theta)$ becomes the joint probability of the consecutive measurement $p(a,b|AB)$. In this case, the coQM is reduced and equivalent to the conventional quantum metrology using the consecutive measurement so that the $\Delta \theta_\mathrm{co}$ equals or larger than $\Delta \theta_q$ (see Supplementary Information). For the contextual model, conversely, $\Delta \theta_\mathrm{co}$ can be smaller than $\Delta \theta_q$ (see Ref.~\cite{Jae2023} for rigorous discussions). Fig.~\ref{fig:result} {\bf a} shows simulation results of the contextuality-enabled enhancement on the Bloch sphere.

We perform polarization estimation of $\theta$ with probe states prepared in $\ket{\psi} = \cos(\theta/2) \ket{H} + e^{i\phi}\sin(\theta/2)\ket{V}$ for $0.46\pi\le\theta\le0.55\pi$ and $\phi=0.15\pi$. For the probe states, the operational quasiprobability is given by $w(a,b|\theta)=\left( 1 + (-1)^a \cos \theta + ( -1)^b \sin \theta \cos \phi \right)/4$. We draw $N_s=10^5$ samples for each data set to construct the operational quasiprobability, and calculate an estimate with the estimator~\eqref{eq:mlew}. For polarization estimation of $\theta$, QFI $F_q=1$, so the coQM gains the contextuality-enabled enhancement if $\Delta \theta_\mathrm{co} < \Delta \theta_q/\sqrt{2} = 1/\sqrt{2N_s}\approx 2.24\times10^{-3}$. The errors of the coQM are smaller than the error limit of the conventional quantum metrology for the whole selected range of $\theta$ (Fig.~\ref{fig:result} $\bf b$). The worst case in our results has $ \Delta\theta_\mathrm{co} \approx 1.53\times10^{-3}$ around $\theta=\pi/2$, and the best case has $\Delta\theta_\mathrm{co} \approx 3.7\times10^{-4}$ around each end of the range of $\theta$. This demonstrates that our method elevates the precision of polarimetry by factor of $1.4$ to $6.0$ from the limit of conventional quantum metrology.

We estimate sucrose solutions of three different concentrations $c=0.1$, $0.3$ and $0.5$ g ml$^{-1}$. We prepare the probe state with initial parameters $\theta_0 = 0.5\pi$ and $\phi=0.15\pi$. For each concentration, we repeat the estimation $10$ times. The results (Fig.~\ref{fig:result} {\bf d}) show that the errors of estimates by the coQM, $\Delta c_\mathrm{co}$, are smaller than the minimum error by the conventional quantum metrology ($\approx5.9\times 10^{-2}$); For $c=0.1$, $0.3$ and $0.5$ g ml$^{-1}$, the mean errors are $\approx3.7\times 10^{-2}$, $\approx3.3\times 10^{-2}$, and $\approx2.8\times 10^{-2}$, respectively. This illustrates that the coQM exceeds the conventional quantum metrology by a wide margin.

%\section*{Discussion and outlook}
\section*{Measurement Incompatibility}
The contextuality of measurement selection stems from the incompatibility of quantum measurements~\cite{Busch86}. In the scenario of the consecutive measurement, if the two measurements $A$ and $B$ commute, the consecutive measurement is de facto a single measurement; the probabilities $p(b|B)$ and $p(b|AB)$ are equal and the prediction for $B$ is noncontextual. Otherwise, the prediction for the measurement $B$ depends on whether performing the first measurement $A$ and it is contextual except for a case when the initial state is prepared in an eigenstate of the $A$. Thus, the measurement incompatibility is necessary for the contextuality of measurement selection.

The noncommutation of observable operators defines the incompatibility among measurements, represented by projection-valued measures (PVM), which we assume in the present work. The notion of incompatibility needs to be generalized if the representation of measurement is generalized to positive operator-valued measure (POVM). This generalization is required, for example, if one considers an open quantum system in a noisy environment. Non-joint measurability (non-JM) is one of the generalizations~\cite{Busch86}. The non-JM plays an important role in a contextuality~\cite{Tavakoli2020,Guhne2023}, as does the noncommutativity~\cite{budroni2021}. In fact, non-JM and the contextuality of measurement selection are also closely related as the negativity of operational quasiprobability is the necessary and sufficient condition for non-JM~\cite{Jae2019,Jae2023}.

Recently, there were studies in a similar vein to the present work~\cite{Arvidsson2020,Noah2022}, showing that the noncommutativity can be a resource for quantum metrology. However, their schemes employ a post-selection to discard unwanted measurement outcomes, so there is a tradeoff between success probability and Fisher information; success probability becomes small if Fisher information is large~\cite{Combes2014}. Quantum post-selected metrology such as weak value amplification methods share this matter~\cite{Ferrie2014,Knee2014}. On the contrary, our method utilizes all of measurement outcomes for the estimation~\cite{Ryu2013,Jae2017}, implying that the coQM is free from such tradeoff.

\section*{Outlook}
This work demonstrates that utilizing the contextuality of measurement selection can enhance the precision measurement. The experiment attains precision beyond the limit of conventional quantum metrology~\cite{Holevo2011,Braunstein1994}. The coQM has advantages over the conventional method (see Supplementary Information): it can enhance the precision without optimizing the measurements if they are incompatible, and it works even without any entangled state of probe that has been regarded difficult to generate and manipulate. We use the heralded single-photon source to clearly show the performance of coQM per a unit particle of probe. We expect that a multi-photon source can also be adopted for the coQM with the similar settings of measurements~\cite{Ryu_2019}. Our method is expected to be applicable to a quantum sensor~\cite{Degen2017} if the context of measurement selection can be implemented within the sensor's system. In addition, the approaches employed to demonstrate the contextuality-enabled enhancements can be utilized to characterize quantum devices (see Supplementary Information), which is a fundamental task required to implement quantum technologies.

%%%%%%%%%%%%%%%%%%%%%%%%%%%%%%%%%%%%%%%%%%%%%%%%%%%%%%%%%%%%%%%%%%%%%%%%%%%%%%%%%%%%

%\bibliography{references}% Produces the bibliography via BibTeX.

%%%%%%%%%%%%%%%%%%%%%%%%%%%%%%%%%%%%%%%%%%%%%%%%%%%%%%%%%%%%%%%%%%%%%%%%%%%%%%%%%%%%

%-----------------------------------------------------------------------------------------------------------------------------------------------------------------------------------------------------------------------------------------------------------------------------------------------------
%-----------------------------------------------------------------------------------------------------------------------------------------------------------------------------------------------------------------------------------------------------------------------------------------------------
\setcounter{equation}{0}
\setcounter{section}{0}
\renewcommand{\d}[1]{\ensuremath{\operatorname{d}\!{#1}}}
\renewcommand{\thesubsection}{M\arabic{subsection}}
\renewcommand{\theequation}{M\arabic{equation}}

%-----------------------------------------------------------------------------------------------------------------------------------------------------------------------------------------------------------------------------------------------------------------------------------------------------
%-----------------------------------------------------------------------------------------------------------------------------------------------------------------------------------------------------------------------------------------------------------------------------------------------------
\section*{Methods}

\subsection{Heralded single photons}
\label{sec:photon}
We generate the heralded single photon as following. High energy pump photons ($p = 405.7$ nm) from a continuous wave (CW) single mode laser (TOPMODE $405$, TOPTICA) are sent to a periodically poled KTP (PPKTP) crystal. PPKTP splits the input photons into photon pairs (signal and idler photons) through type-II spontaneous parametric down conversion (SPDC) process. The polarizations of signal and idler photons are orthogonal each other, so that polarizing beam splitter (PBS) can separate them into two different optical paths. The idler photon is sent to an avalanche photodiode (APD) for triggering. The signal photon is sent to one of the four APDs (SPCM-QC$4$, Perkin Elmer). If the trigger APD is clicked, we count clicks on the four APDs. We control the count rate of the trigger APD to be $~ 2\times10^5$ cps to sufficiently suppress multi photon events, i.e., $\ket{\text{SPDC}} \approx \ket{00} + \epsilon\ket{11}$ for $\epsilon \ll 1$. The click signals are post-processed by a field programmable gate array (FPGA) with the time bin size of $25$ ns.

\subsection{Input state preparation}
\label{sec:preparation}
We prepare an initial probe state by using a series of three wave plates (Fig.~\ref{fig:experiment}), one of half-wave plate (HWP) and two of quarter-wave plates (QWP). After passing QWP$_1$, HWP, and QWP$_2$ sequentially, a horizontally polarized state $\ket{H}$ becomes an initial state
\begin{eqnarray}
\ket{\psi}_{\text{in}} &=& \text{QWP}_2\left( \frac{\pi}{4} \right) \text{HWP}\left(p \right)\text{QWP}_1\left( q \right) \ket{H} \nonumber\\
&=& e^{i \left(-2p + q + \pi/4 \right)}
\begin{pmatrix}
\cos\left(\frac{\pi}{4} - q \right) \\
e^{i \left(4p - 2q - \frac{\pi}{2} \right)} \sin\left(\frac{\pi}{4} - q \right)
\end{pmatrix},
\end{eqnarray}
where $p$ $(q)$ is the angle of the fast axis of the half (quarter)-wave plate from the horizontal axis. The $q$ value of QWP$_2$ is fixed at $\pi/4$. By adjusting the control parameters $p$ and $q$ to satisfy $\theta = \pi/2 - 2q$ and $\phi = 4p - 2q - \pi/2$, we finally obtain the parameterized state $\ket{\psi}_\text{in} = \cos(\theta/2) \ket{H} + e^{i\phi} \sin(\theta/2) \ket{V}$.

\subsection{Maximum likelihood estimator using operational quasiprobability}
\label{sec:mleoq}

Maximum likelihood estimation is a method to find a parameter of a probability model which best describes observed data. This method assumes a likelihood function of the model, and maximizes the function to determine the most likely value in the parameter space as an estimate. In this work, we take the operational quasiprobability as the model depending on the phase $\theta$.

For the two data sets $\mathbf{x}_B$ and $\mathbf{x}_{AB}$, we define the log-likelihood function as
\begin{align}
l_W(\theta|\mathbf{x}_B,\mathbf{x}_{AB}) 
&:= \frac{1}{N_s} \sum_{a,b=0}^1 N_W(a,b)\log w(a,b|\theta),
\end{align}
where $N_W(a,b) = N_{AB}(a,b) + \left(N_B(b) - N_{AB}(b) \right)/2$. $N_{B}(b)$ is the number of counts for outcome $b$ in the data set $\mathbf{x}_{B}$, and $N_{AB}(a,b)$ is the number of counts for outcome pair $(a,b)$ in the data set $\mathbf{x}_{AB}$. $N_{AB}(b)$ is obtained by the marginal number of counts as $N_{AB}(b)=\sum_a N_{AB}(a,b)$. For a small number of samples, $N_w(a,b)$ can be negative by statistical fluctuations. We test whether the number count $N_W$ is positive, and neglect cases where the count is negative.

In a broader sense, the coQM proposes an approach of integrating the two different ensembles for single-parameter estimation. To show that our estimator $\check{\theta}$ is unbiased and error of the estimate achieves Cram{\'e}r-Rao bound~\cite{Cramer1946,Rao1992}, we propose a theory that describes the operational quasiprobability as an ensemble mixture model (see Supplementary Information).

\begin{acknowledgements}

MSK acknowledges the EPSRC grant (EP/T00097X/1) and AppQInfo MSCA ITN from the European Unions Horizon 2020. KGL was supported by the National Research Foundation of Korea (NRF) grant funded by the Korea government (MSIT) (No. 2023M3K5A109481311) and Institute of Information and Communications Technology Planning \& Evaluation (IITP) grant funded by the Korea government (MSIT) (No. 2022-0-01026). JL was supported by the National Research Foundation of Korea (NRF) grant funded by the Korea government (MSIT) (No. 2022M3E4A1077369).

\subsection*{Author contributions}
Jeongwoo Jae, M. S. Kim, and Jinhyoung Lee contributed to the theoretical formulation of the contextual quantum metrology. Jiwon Lee and Kwang-Geol Lee conducted the optical experiments and the data analysis. All authors contributed to discussions in this work. Jeongwoo Jae wrote the manuscript with assistance of other authors. Jeongwoo Jae and Jiwon Lee contributed equally to this work.

\subsection*{Competing interests}

The authors declare no competing interests.

\subsection*{Supplementary Information}

Supplementary Information is available for this paper.
\end{acknowledgements}

\clearpage
\onecolumngrid
%-----------------------------------------------------------------------------------------------------------------------------------------------------------------------------------------------------------------------------------------------------------------------------------------------------
%-----------------------------------------------------------------------------------------------------------------------------------------------------------------------------------------------------------------------------------------------------------------------------------------------------
\setcounter{equation}{0}
\setcounter{section}{0}
\setcounter{figure}{0}
\renewcommand{\d}[1]{\ensuremath{\operatorname{d}\!{#1}}}
\renewcommand{\thesection}{S\arabic{section}}
\renewcommand\thesubsubsection{\Alph{subsubsection}}
\renewcommand\thesubsection{\thesection-\arabic{subsection}}
\renewcommand{\theequation}{S\arabic{equation}}
\renewcommand{\thefigure}{S\arabic{figure}}
%-----------------------------------------------------------------------------------------------------------------------------------------------------------------------------------------------------------------------------------------------------------------------------------------------------
%-----------------------------------------------------------------------------------------------------------------------------------------------------------------------------------------------------------------------------------------------------------------------------------------------------

\centerline{\bf\large{{Supplementary Information for contextual quantum metrology}}}

\vspace{0.5cm}
We provide Supplementary Information (SI) including details of the present work: contextual quantum metrology. SI is organized into three parts. First, we briefly review the conventional quantum estimation theory. Second, we recall asymptotic normality of maximum likelihood estimator (MLE), and suggest ensemble mixing theory to prove that MLE using operational quasiprobability has the asymptotic normality. Based on the proof, we suggest an estimator of the estimation error. By using this estimator, we experimentally show that there is the contextuality-enabled enhancement also for estimation of the phase $\phi$. Third, we characterize systematic errors, and investigate effects of noise caused by depolarization of probe state on the performance of contextual quantum metrology.

\section{quantum estimation theory}

%\subsection{Quantum estimation theory}
Estimation or metrology theory employs two models. One is a conditional probability function $p({\mathbf x} | \theta)$ of an observing data set $\mathbf{x}$, conditioned on parameter $\theta$ to estimate. The other is an estimator $T({\mathbf x})$, which estimates the value of parameter $\theta$, based on the observation. Both types of models are verified whether appropriate for experiments. To find the two types of models with small estimation errors is at the heart of estimation. Once a conditional probability function is assumed, the estimator determines the estimation error.

The estimation error $\Delta \theta$ is known to have a lower bound over all estimators, 
\begin{align}
\label{eq:crb}
\Delta \theta \ge \frac{1}{\sqrt{N_s F}}.
\end{align}
This inequality holds when samples are drawn from i.i.d. The lower bound is called Cram{\'e}r-Rao bound (CRB). CRB contains Fisher information $F$, which is defined by the conditional probability and its dependence of parameter $\theta$:
\begin{align}
F = F(\theta) = \sum_x p(x|\theta) \left( \frac{\partial \ln p (x|\theta)}{\partial \theta} \right)^2. 
\end{align}
Note that Fisher information is positive semidefinite. If the conditional probability is parameter-independent, $F$ has the minimum of zero, $F=0$. If it is highly dependent, $F$ is large. As $F$ increases, CRB decreases and results in more precise estimation. An estimator attaining CRB is said optimal. It is known that a maximum likelihood estimator is (approximately) an optimal estimator in limit of large samples~\cite{lehmann2006}.

In quantum theory, the conditional probability $p(x|\theta)$ is decomposed into a quantum state $\hat{\varrho}$ and a measurement of positive operator-valued measure (POVM) $\{\hat{m}(x)\}$. Assuming the quantum state is a functional of parameter $\theta$, the conditional probability is given by $p(x|\theta) = \text{Tr}\left( \hat{m}(x) \hat{\varrho}(\theta) \right)$. In this case, CRB can be further optimized over measurements to quantum CRB (QCRB). QCRB contains quantum Fisher information (QFI) defined by
\begin{equation}
    F_q = \text{Tr} (\hat{L}^2 \hat{\varrho}(\theta)),
\end{equation}
where $\hat{L}:=({\hat{L}\hat{\varrho}(\theta)+\hat{\varrho}(\theta)\hat{L}})/{2}$ is the symmetric logarithmic operator to define derivative of a quantum state $\hat{\varrho}(\theta)$~\cite{Braunstein1994}. For a given probe state, QFI is the maximum value of Fisher information over all measurements, so it is never less than Fisher information $F$, i.e., $F_q \ge F$. Equivalently, the error given by quantum Fisher information is never larger than the error given by Fisher information. In that sense, the error given by QFI is the minimum error of quantum metrology using measurements which can be represented by a POVM.

\section{Maximum likelihood estimator using operational quasiprobability}

\subsection{Maximum likelihood estimator}

We recall the significant properties of maximum likelihood estimator (MLE) to show asymptotic normality of MLE using operational quasiprobability (MLEOQ). The MLE estimates a parameter $\theta$ by using a data set ${\bf x}=\{x_1,x_2,\cdots,x_N\}$ drawn from a single ensemble $X$ represented by a probability $p(x|X,\theta)$. An estimate $\check{\theta}$ is a value that gives an extremum of log-likelihood function, which is defined by
\begin{equation}
    l_X(\theta|{\bf x}) := \frac{1}{N} \sum_{x_i\in {\bf x}} \log p(x_i|W,\theta).
\end{equation}
The log-likelihood function satisfies following three conditions in the asymptotic limit of $N\rightarrow\infty$~\cite{lehmann2006}:
\begin{eqnarray}
    (\text{C}.1)&&~~-\partial^2_{\theta} l_X(\theta|{\bf x}) = F_X(\theta) + {\cal O}\left( {N}^{-1/2}\right) \nonumber\\%,~\text{as}~ N\rightarrow\infty. \nonumber\\
    (\text{C}.2)&&~~~{\bf {E}}\left[ \partial_\theta l_X(\theta|{\bf x}) \right] =0. \nonumber \\
    (\text{C}.3)&&~~~{\bf {E}}\left[ \left(\partial_\theta l_X(\theta|{\bf x}) \right)^2 \right] = F_X(\theta)/N, \nonumber
\end{eqnarray}
where $F_X(\theta)$ is the Fisher information of probability $p(x|X,\theta)$.

If we assume that an estimate $\check{\theta}$ is close to a true parameter value $\theta_0$, the difference between the estimate and the true value can approximate to
\begin{eqnarray}
    \check{\theta} - \theta_0 \approx -\frac{\partial_\theta l(\theta_0|{\bf x}) }{\partial^2_\theta l(\theta_0|{\bf x})}.
    %{\partial_\theta l(\theta_0|{\bf x}) }\approx (\theta - \theta_0){\partial^2_\theta l(\theta_0|{\bf x})}
    %{\bf E}\left[{\partial_\theta l(\theta_0|{\bf x}) }\right] &\approx& (\theta - \theta_0){\bf E}\left[{\partial^2_\theta l(\theta_0|{\bf x})}\right]
\end{eqnarray}
The first order moment of the difference determines biasedness, quantifying how much the estimate is shifted from the true value. The second order moment of the difference corresponds to estimation error. 

The conditions (C.$1$) and (C.$2$) imply that the MLE is an unbiased estimator:
\begin{equation}
\label{Seq:unbias}
    {\bf E}\left[\check{\theta} - \theta_0\right] \approx \frac{{\bf E}\left[\partial_\theta l(\theta_0|{\bf x}) \right]}{F_X(\theta_0) + {\cal O}\left({N}^{-1/2}\right)} = 0.
\end{equation}
The (C.$2$) and (C.$3$) imply that MLE is an efficient estimator; the estimation error $\Delta^2\theta$ attains Cram{\'e}r-Rao bound in the asymptotic limit of $N\rightarrow \infty$,
\begin{equation}
\label{Seq:efficiency}
    \Delta^2\theta = {\bf E}\left[\left(\check{\theta} - \theta_0\right)^2\right] \approx \frac{{\bf E}\left[\left(\partial_\theta l(\theta_0|{\bf x}) \right)^2\right]}{\left[F_X(\theta_0) + {\cal O}\left({N}^{-1/2}\right)\right]^2} = \frac{1}{NF_X(\theta_0)} + {\cal O}(N^{-3/2}).
\end{equation}
The unbiasedness and the efficiency finally imply that MLE satisfies asymptotic normality in the limit $N\rightarrow \infty$, i.e.,
\begin{equation}
    \sqrt{N}\left(\check{\theta} - \theta_0 \right) \xrightarrow{N} {\cal N}(0,F^{-1}_X(\theta_0)),
\end{equation}
where $\cal N$ is normal distribution of zero mean and variance $F^{-1}_X(\theta_0)$. Purpose of following sections is to prove that the MLEOQ has the asymptotic noramlity.

\subsection{Log-likelihood function of operational quasiprobability}

Whereas the typical MLE is based on a single ensemble $X$ represented by a conditional probability $p(x|X,\theta)$, MLE in the contextual quantum metrology (coQM) utilizes the two ensembles $B$ and $AB$ represented by the probabilities $p(b|B)$ and $p(a,b|AB)$, respectively. To represent the probabilities from these two ensembles at once, we employ the operational quasiprobability (OQ) given by
\begin{equation}
\label{Seq:OQ}
    w(a,b|W,\theta) = p(a,b|AB,\theta) + \frac{1}{2}\left(p(b|B,\theta) - p(b|AB,\theta) \right).
\end{equation}
One can directly collect samples from the two ensembles $B$ and $AB$ through the experiment shown in Fig.~$1$, but the samples of $W$ can only be deduced from those of $B$ and $AB$ as a mixture of two ensembles. In this sense, we call the ensembles $B$ and $AB$ observable, and $W$ virtual.

Our approach uses the virtual ensemble $W$ to estimate the parameter $\theta$. MLE using OQ (MLEOQ) calculates an estimate as a value which gives an extremum of log-likelihood function of $w$ function, which is defined by
\begin{equation}
\label{Seq:logOQ}
    l_W(\theta|{\bf x}_W) := \frac{1}{N_W} \sum_{x_i\in {\bf x}_W} \log w(x_i|W,\theta),
\end{equation}
where ${\bf x}_W$ is a composite data set of ${\bf x}_{B}$ and ${\bf x}_{AB}$ drawn from the observable ensembles $A$ and $AB$, respectively. Note that we only consider positive OQ to apply the function $w$ to the logarithmic function without divergence. We test whether the OQ is positive or not with negativity defined by $\sum_{a,b} \left( \abs{w(a,b|W,\theta)} - w(a,b|W,\theta) \right)/2$. $w$ is positive semidefinite if the negativity is zero.

A lot of work in quantum metrology have been done based on the theory of typical maximum likelihood estimation assuming a single observable ensemble. On the other hand, estimation based on an ensemble mixture has been drawn little attention in the field of quantum metrology. In this work, we propose a maximum likelihood estimation utilizing the two different ensembles. To do so, we suggest a theory of ensemble mixing and prove that MLEOQ satisfies the asymptotic normality.

\begin{figure*}[b!]
    \centering
    \includegraphics[width=0.75\textwidth]{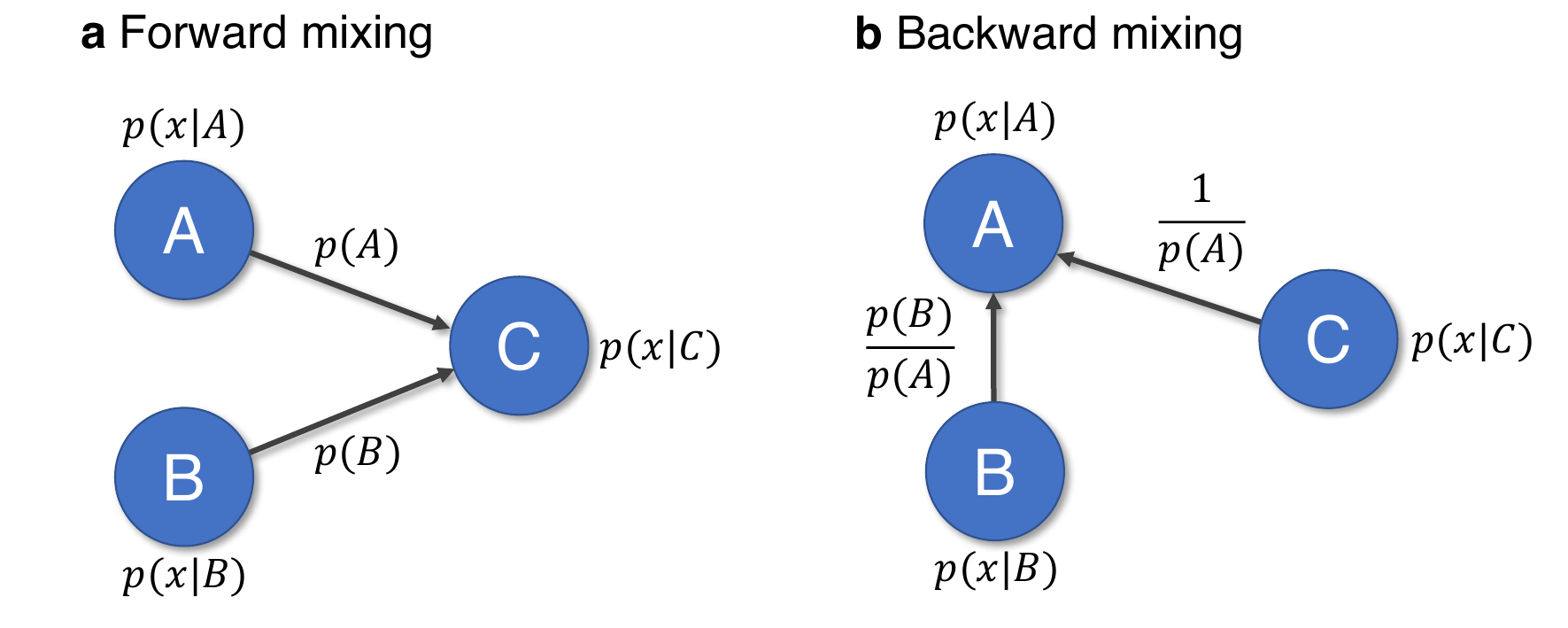}
    \caption{{\bf Diagram of forward and backward mixing of ensembles} }
    \label{figS:mixing}
\end{figure*}

\subsection{Forward mixing of ensembles}

Suppose that a virtual ensemble $C$ is a mixture of two observable ensembles $A$ and $B$ by following relation
\begin{equation}
\label{Seq:mixing}
    p(x|C) = p(x|A)p(A) + p(x|B)p(B),
\end{equation}
where $p(x|X)$ is a probability which represents a ensemble $X$, and $p(X)$ is a probability to choose ensemble $X$. If one draws $N_A$ samples from the ensemble $A$ and $N_B$ from $B$, the number of samples of $C$ is $N_C=N_A + N_B$. Thus, in the asymptotic limit of $N_C\rightarrow \infty$, $p(A)=N_A/N_C$ and $P(B)=N_B/N_C$. We describe the virtual ensemble $C$ as a statistical mixture of two observable ensembles $A$ and $B$. We call this description as forward mixing of ensembles. The forward mixing is depicted in Fig.~\ref{figS:mixing}. {\bf a}.

We call one of significant properties of the forward mixing as {\em functional equivalence}: expectation of an arbitrary function $f$ by the virtual ensemble $C$ can be decomposed into those by the observable ensembles $A$ and $B$. In other words,
\begin{equation}
\label{Seq:equivalence}
    \E{f|C} = p(A)\E{f|A} + p(B)\E{f|B},
\end{equation} 
where $\E{f|X}=\sum_{x} f(x)p(x|X)$. This is a key property to show the asymptotic normality of MLEOQ.

To prove the functional equivalence, we analyze the fluctuation of number count $N(x|X)$ for an outcome $x$ of a data set ${\bf x} = \{ x_1,x_2,\cdots,x_{N_X} \}$ drawn from an ensemble $X$. For each outcome $x$, the number of count $N(x|X)$ can approximate to mean value $\bar{N}(x|X):=N_Xp(x|X)$ with standard deviations $\sigma=\sqrt{N_X}$ in the asymptotic limit $N_X\rightarrow\infty$ as
\begin{equation}
\label{Seq:Nfluctuation}
    N(x|X) \approx \bar{N}(x|X) + {\cal O}\left( \sqrt{N_X} \right).
\end{equation}

Let a grand probability distribution of count numbers be $p(\{N(x|X)\}) = \Pi_x p(N(x|X))$. Each probability $p(N(x|X))$ respective to a number count follows a binomial distribution $\binom{N}{N(x|X)}p(x|X)^{N(x|X)}(1-p(x|X))^{N-N(x|X)}$. In the asymptotic limit of $N_X \rightarrow \infty$, the Binomial distribution approximates to a Poisson distribution of mean value $\bar{N}(x|X)$. Furthermore, as the average number count increases along with the increase of $N_X$, the Poisson distribution approximates to a normal distribution ${\cal N}(\bar{N}(x|X),\bar{N}(x|X))$. Thus, the grand probability finally approximates to the normal distribution as
\begin{eqnarray}
    p(\{N(x|X)\}) \propto \exp\left[-\frac{1}{2}\sum_x \left(\frac{N(x|X) - \bar{N}(x|X)}{\bar{N}(x|X)}\right)^2\right]. \nonumber
\end{eqnarray}
This implies that $N(x|X)\approx \bar{N}(x|X) \pm \sqrt{N_Xp(x|X)}$ in the asymptotic limit of $N_X\rightarrow\infty$. %\hfill$\blacksquare$

By using the approximation of number count~\eqref{Seq:Nfluctuation}, we show that the number count of virtual ensemble $C$ is decomposed of those of $A$ and $B$ as
\begin{equation}
\label{Seq:rel2}
    N(x|C) \approx N(x|A) + N(x|B) + {\cal O}\left( \sqrt{N_{A,B,C}}\right).
\end{equation}
From the relation~\eqref{Seq:mixing}, we can derive that
\begin{eqnarray}
    N(x|C) &=& N_C p(x|C) \nonumber \\
    &=& N_C p(x|A)p(A) + N_C p(x|B)p(B) \nonumber\\
               &\approx& {N_A} p(x|A) + N_B p(x|B) + {\cal O}\left( \sqrt{N_{C}} \right) \nonumber \\
               &\approx& N(x|A) + N(x|B) + {\cal O}\left( \sqrt{N_{A,B,C}} \right), \nonumber
\end{eqnarray}
where we use $N_X \approx N_C p(X) + {\cal O}\left(\sqrt{N_C}\right)$ for the third step and $N(x|X) \approx N_X p(x|X) + {\cal O}\left( \sqrt{N_X} \right)$ for the fourth step. %\hfill$\blacksquare$

Finally, using the relation~\eqref{Seq:rel2}, we prove the functional equivalence. For the data set of ensemble $C$, ${\bf x_C}$, the expectation of function $f$ can be read
\begin{eqnarray}
\label{Seq:functionaleq}
    {\bf E}\left[{\sum_{x_i\in {\bf x}_C} f(x_i)}\right]
    %&=& N_C{\bf E}\left[f|C\right]  \nonumber \\
    &=& {\bf E}\left[\sum_{x} N(x|C)f(x)\right] \approx N_C{\bf E}\left[f|C\right] + {\cal O}\left( \sqrt{N_C}
    \right)\nonumber\\
    &\approx& {\bf E}\left[{\sum_{x} N(x|A)f(x)}\right] + {\bf E}\left[{\sum_{x} N(x|B)f(x)}\right] + {\cal O}\left(\sqrt{N_{A,B,C}}\right) \nonumber \\
    &=& {\bf E}\left[{\sum_{x_j\in {\bf x}_A} f(x_j)}\right] + {\bf E}\left[{\sum_{x_k\in {\bf x}_B} f(x_k)}\right] + {\cal O}\left(\sqrt{N_{A,B,C}}\right) \nonumber \\
    &=& N_A{\bf E}\left[f|A\right] + N_B{\bf E}\left[f|B\right] + {\cal O}\left(\sqrt{N_{A,B,C}}\right).
\end{eqnarray}
This implies that $N_C{\bf E}\left[f|C\right]  = N_A{\bf E}\left[f|A\right] + N_B{\bf E}\left[f|B\right]$ or, equivalently, ${\bf E}\left[f|C\right]  = p(A){\bf E}\left[f|A\right] + p(B){\bf E}\left[f|B\right]$ in the asymptotic limit of $N_{A,B,C} \rightarrow \infty$. \hfill$\blacksquare$

\subsection{Backward mixing of ensembles}
We can describe the forward mixing in a different view that considers the observable ensemble $A$ as a mixture of observable ensemble $B$ and virtual ensemble $C$. We call this description as backward mixing of ensembles (see Fig.~\ref{figS:mixing}. {\bf b}). In the backward mixing, we have a reciprocal relation of Eq.~\eqref{Seq:mixing},
\begin{equation}
    p(x|A) = p(x|C)\frac{1}{p(A)} - p(x|B)\frac{p(B)}{p(A)}.
\end{equation}
The backward mixing preserves the functional equivalence, i.e.,
\begin{equation}
\label{Seq:back_equivalence}
    \E{f|A} = \E{f|C}\frac{1}{p(A)} - \E{f|B}\frac{p(B)}{p(A)}.
\end{equation} 
In the backward problem, the number count satisfies a reciprocal relation of number counts of Eq.~\eqref{Seq:rel2}.
\begin{eqnarray}
\label{Seq:rel2back}
    N_A p(x|A) &=& N_A p(x|C)\frac{1}{p(A)} - N_A p(x|B)\frac{p(B)}{p(A)} \nonumber\\
               &\approx& \frac{N_A}{N_C} {N_C}p(x|C)\frac{1}{p(A)} - \frac{N_A}{N_B} N_B p(x|B)\frac{p(B)}{p(A)} + {\cal O}\left( \sqrt{N_{A}} \right) \nonumber \\
               &\approx& N(x|C) - N(x|B) + {\cal O}\left( \sqrt{N_{A,B,C}} \right). 
\end{eqnarray}
Thus, the functional equivalence becomes
\begin{eqnarray}
\label{Seq:functionaleq_back}
    {\bf E}\left[{\sum_{x_i\in {\bf x}_A} f(x_i)}\right]
    %&=& N_A{\bf E}\left[f|A\right]  \nonumber \\
    &=& {\bf E}\left[\sum_{x} N(x|A)f(x)\right] \approx N_A{\bf E}\left[f|A\right] + {\cal O}\left(\sqrt{N_A}\right) \nonumber\\
    &\approx& {\bf E}\left[{\sum_{x} N(x|C)f(x)}\right] - {\bf E}\left[{\sum_{x} N(x|B)f(x)}\right] + {\cal O}\left(\sqrt{N_{A,B,C}}\right) \nonumber \\
    &=& {\bf E}\left[{\sum_{j\in C} f(x_j)}\right] + {\bf E}\left[{\sum_{k\in B} f(x_k)}\right] + {\cal O}\left(\sqrt{N_{A,B,C}}\right) \nonumber \\
    &=& N_C{\bf E}\left[f|C\right] - N_B{\bf E}\left[f|B\right] + {\cal O}\left(\sqrt{N_{A,B,C}}\right).
\end{eqnarray}
This implies that $N_A{\bf E}\left[f|A\right]  = N_C{\bf E}\left[f|C\right] - N_B{\bf E}\left[f|B\right]$ and, equivalently, ${\bf E}\left[f|A\right]  = {\bf E}\left[f|C\right]\frac{1}{p(A)} - {\bf E}\left[f|B\right]\frac{p(B)}{p(A)}$ in the asymptotic limit of $N_{A,B,C} \rightarrow \infty$. \hfill$\blacksquare$

\begin{figure*}[b!]
    \centering
    \includegraphics[width=0.75\textwidth]{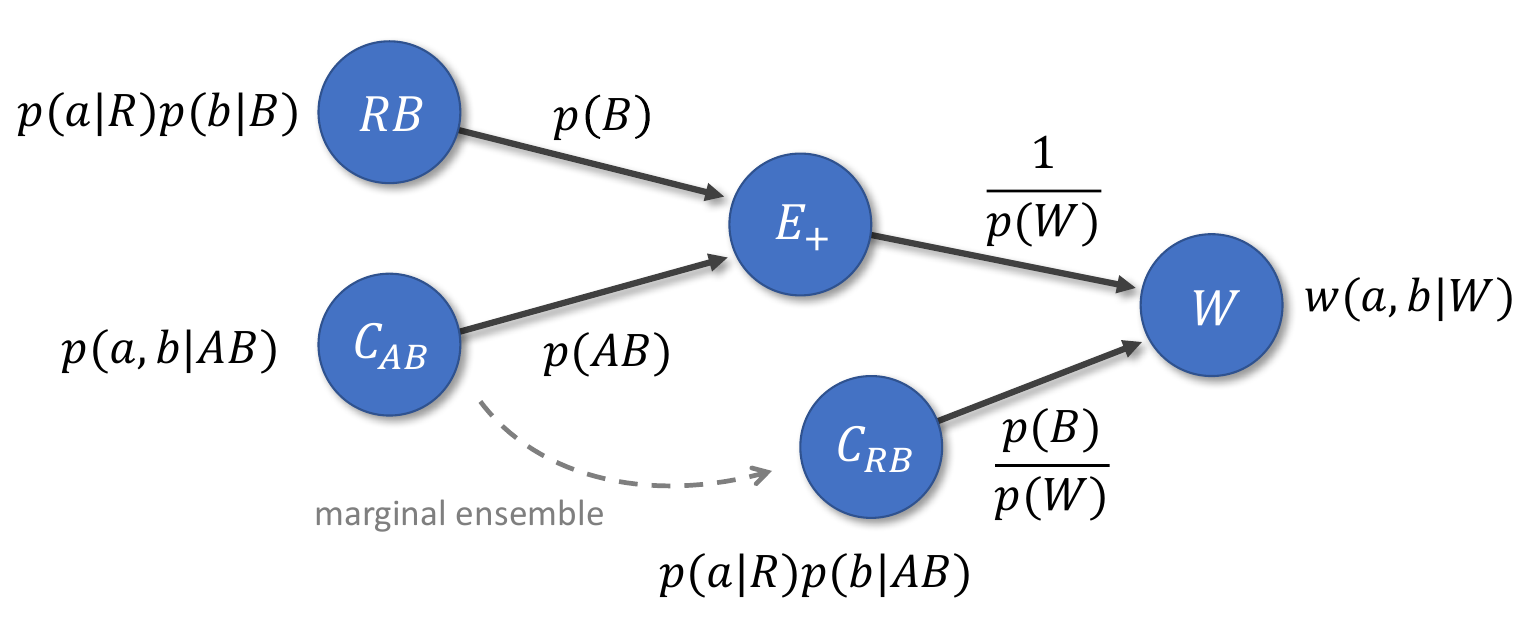}
    \caption{{\bf Diagram of ensemble mixing to construct operational quasiprobability} }
    \label{figS:OQmixing}
\end{figure*}

\subsection{Operational quasiprobability as a distribution of ensemble mixtures}
Now we are in a position to interpret the operational quasiprobability (OQ) as a model to represent the forward and backward mixture of ensembles. For the OQ, observable ensembles are $RB$, $C_{AB}$, and $C_{RB}$, where $R$ is a random ensemble, $B$ is the ensemble for the single $B$ measurement, $C_{AB}$ is the ensemble for the consecutive measurement $AB$, and $C_{RB}$ is the marginal ensemble $B$ of $C_{AB}$. 

The OQ is obtained by the backward mixing of a virtual ensemble, called $E_+$, and the marginal ensemble $C_{RB}$ as
\begin{eqnarray}
    w(a,b|W) &=& p(a,b|E_+)\frac{1}{p(W)} - p(a|R)p(b|AB)\frac{p(B)}{p(W)}, %\\
    %p(a,b|E_+) &=& p(a|R)p(b|B)p(B) + p(a,b|AB)p(AB),
\end{eqnarray}
where the virtual ensemble $E_+$ is defined by the forward mixing of observable ensembles $RB$ and $C_{AB}$ as
\begin{eqnarray}
    p(a,b|E_+) = p(a|R)p(b|B)p(B) + p(a,b|AB)p(AB).
\end{eqnarray}
We here consider $p(W)=1/2$, $p(B)=1/2$, $p(AB)=1/2$, and $p(a|R)=1/2$ $\forall a$, and the OQ in Eq.~\eqref{Seq:OQ} is obtained by
\begin{eqnarray}
    w(a,b|W) 
    &=& \frac{p(AB)}{p(W)}p(a,b|AB) + \frac{p(B)}{p(W)}p(a|R)p(b|B) - \frac{p(B)}{p(W)}p(a|R)p(b|AB)\nonumber \\
    &=& p(a,b|AB) + \frac{1}{2}p(b|B) - \frac{1}{2}p(b|AB).
\end{eqnarray}
Diagram for the ensemble mixings to construct the OQ is presented in Fig.~\ref{figS:OQmixing}

\subsection{Asymptotic normality of MLEOQ}
Based on the aforementioned asymptotic properties, we show that the log-likelihood function of OQ satisfies following three conditions:
\begin{eqnarray}
    (\text{C}'.1)&&~~-\partial^2_{\theta} l_W(\theta|{\bf x}) = F_W(\theta) + {\cal O}\left( {N}^{-1/2}_W\right),~\text{as}~ N_W\rightarrow\infty. \nonumber\\
    (\text{C}'.2)&&~~~{\bf {E}}\left[ \partial_\theta l_W(\theta|{\bf x}_W) \right] =0. \nonumber \\
    (\text{C}'.3)&&~~~{\bf {E}}\left[ \left(\partial_\theta l_W(\theta|{\bf x}_W) \right)^2 \right] = F_W(\theta)/{N_W }, \nonumber
\end{eqnarray}
Note that we consider a positive $w$ function.

Proof of $(\text{C}'.1)$.---In the asymptotic limit of $N_W\rightarrow\infty$, the second order derivative of log-likelihood function of OQ can be read
\begin{eqnarray}
    -\partial^2_\theta l_W(\theta|{\bf x}_W) &=& -\frac{1}{N_W} \sum_{x_i\in{\bf x}_W} \partial^2_\theta\log w(x_i|W,\theta) \nonumber \\
    &=& -\frac{1}{N_W} \sum_{x} N(x|W) \partial^2_\theta\log w(x|W,\theta) \nonumber \\
    &\approx& -\frac{1}{N_W} \sum_{x} \left[N_W w(x|W,\theta)  + {\cal O}\left(\sqrt{N_W}\right) \right] \left[\frac{\partial^2_\theta w(x|W,\theta)}{w(x|W,\theta)} - \left(\partial_\theta\log w(x|W,\theta)\right)^2 \right]\nonumber \\%\partial^2_\theta\log w(x|W,\theta) 
    &=& \sum_{x} w(x|W,\theta)  \left( {\partial_\theta \log w(x|W,\theta)} \right)^2 +  {\cal O}\left( {N_W}^{-1/2}\right) \nonumber \\
    &=& F_W(\theta) +  {\cal O}\left( {N_W}^{-1/2}\right), \nonumber
\end{eqnarray}
where Fisher information of $w$ is defined by
$F_W(\theta) := \sum_{x} w(x|W,\theta) \left( {\partial_\theta \log w(x|W,\theta)} \right)^2$. \hfill$\blacksquare$

Proof of $(\text{C}'.2)$.---By the functional equivalence in Eq.~\eqref{Seq:functionaleq} and Eq.~\eqref{Seq:functionaleq_back}, $ {\bf E}\left[\partial_\theta l_W(\theta|{\bf x}_W)\right]$ can be decomposed into the respective expectations of observable ensembles $B$ and $AB$ in the asymptotic limit of $N_W \rightarrow \infty$ as
\begin{eqnarray}
    {\bf E}\left[\partial_\theta l_W(\theta|{\bf x}_W)\right] &=& {\bf E}\left[\partial_\theta l_{AB}(\theta|{\bf x}_{AB})\right] + \frac{1}{2}{\bf E}\left[\partial_\theta l_{B}(\theta|{\bf x}_{B})\right] - \frac{1}{2}{\bf E}\left[\partial_\theta l_{B}(\theta|{\bf x}_{AB})\right]. \nonumber
\end{eqnarray}
By the condition $(\text{C}.2)$, each term in the right-hand side is zero. Thus, ${\bf E}\left[\partial_\theta l_W(\theta|{\bf x}_W)\right]$ is zero. \hfill$\blacksquare$

Proof of $(\text{C}'.3)$.--- We use the condition $(\text{C}'.2)$ to prove the condition $(\text{C}'.3)$:
\begin{eqnarray}
{\bf {E}}\left[ \left(\partial_\theta l_W(\theta|{\bf x}_W) \right)^2 \right] &=& \frac{1}{N_W^2} \mathbf{E} \left[ \sum_{x_i,x_j \in {\bf x}_W} \partial_\theta \log w(x_i|W,\theta) \partial_\theta \log w(x_j|W,\theta) \right] \nonumber\\
&\approx& \frac{1}{N_W^2} \sum_{i=j} \mathbf{E} \left[\left( \partial_\theta \log w(x_i|W,\theta) \right)^2 \right]  + \frac{1}{N_W^2} \sum_{i \ne j} \mathbf{E} \left[\partial_\theta \log w(x_i|W,\theta) \right] \mathbf{E} \left[\partial_\theta \log w(x_j|W,\theta)\right]  \nonumber \\
&=& \frac{1}{N_W} \mathbf{E} \left[\left( \partial_\theta \log w(x|W,\theta) \right)^2 \right]\nonumber \\
&=& \frac{1}{N_W} \sum_{x} w(x|W,\theta)\left( \partial_\theta \log w(x|W,\theta) \right)^2 \nonumber \\
&=& {F_W(\theta)}/{N_W} \nonumber~~~~~~~~~~~~~~~~~~~~~~~~~~~~~~~~~~~~~~~~~~~~~~~~~~~~~~~~~~~~~~~~~~~~~~~~~~~~~~~~~~~~~~~~~~~~~~~~~~~~~~~~~~~~~~~\blacksquare
\end{eqnarray}

The conditions $(\text{C}'.1)$ and $(\text{C}'.2)$ imiply that the MLEOQ is unbiased estimator:
\begin{equation}
%\label{Seq:unbias}
    {\bf E}\left[\check{\theta} - \theta_0\right] \approx \frac{{\bf E}\left[\partial_\theta l_W(\theta_0|{\bf x}_W) \right]}{F_W(\theta_0) + {\cal O}\left({N_W}^{-1/2}\right)} = 0.
\end{equation}
The $(\text{C}'.2)$ and $(\text{C}'.3)$ imply that MLEOQ is an efficient estimator as
\begin{equation}
%\label{Seq:efficiency}
    \Delta^2\theta = {\bf E}\left[\left(\check{\theta} - \theta_0\right)^2\right] \approx \frac{{\bf E}\left[\left(\partial_\theta l_W(\theta_0|{\bf x}_W) \right)^2\right]}{\left[F_W(\theta_0) + {\cal O}\left({N_W}^{-1/2}\right)\right]^2} = \frac{1}{N_W F_W(\theta_0)} + {\cal O}(N_W^{-3/2}).
\end{equation}
Thus, the MLEOQ satisfies the asymptotic normality, i.e.,
\begin{equation}
    \sqrt{N_W}\left(\check{\theta} - \theta_0 \right) \xrightarrow{N_W} {\cal N}(0,F_W^{-1}(\theta_0)).
\end{equation}

\begin{figure*}[t!]
    \centering
    \includegraphics[width=\textwidth]{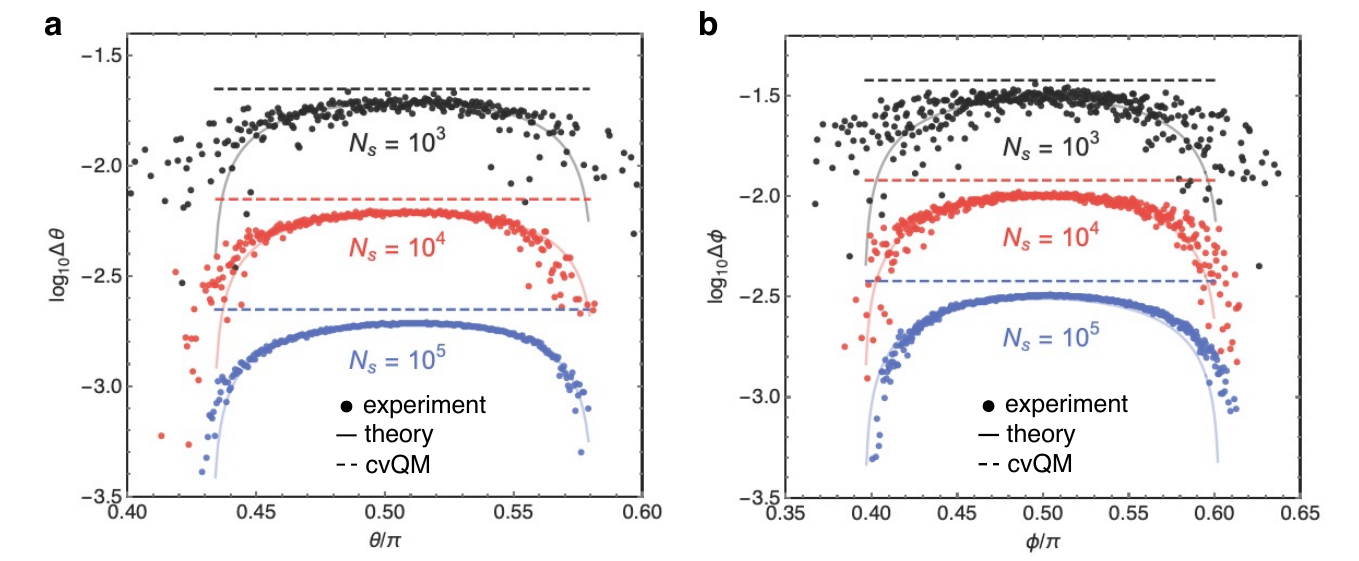}
    \caption{{\bf Contextuality-enabled precision enhancement for $\theta$ and $\phi$ estimation}. We draw $N_s = 10^3$, $10^4$ and $10^5$ samples and, for all cases, we can attain the contextuality-enabled enhancement. For $\theta$ estimation, we prepare probe states $\ket{\psi} = \cos\theta/2\ket{H} + e^{i\phi}\sin\theta/2\ket{V}$ for $\theta \in [0.4\pi,0.6\pi]$ and $\phi=0.2\pi$. For $\phi$ estimation, probe states are $\ket{\psi} = \cos\theta/2\ket{H} + e^{i\phi}\sin\theta/2\ket{V}$ for $\theta=0.2\pi$ and $\phi\in[0.35\pi,0.65\pi]$. The dashed lines are the minimum error in conventional quantum metrology (cvQM) determined by quantum Fisher information.}
    \label{sfig:sample}
\end{figure*}

Based on these facts, to estimate the estimation error ${\bf E}\left[\left(\check{\theta} - \theta_0\right)^2\right]$ in the asymptotic limit of $N_W\rightarrow\infty$, we employ an estimator for Fisher information~\cite{Efron1978,Bruce1997} defined by
\begin{align}
\label{Seq:fie}
{F}_\text{est}({\theta}_\text{est} | \mathbf{x}) &:= \left.- \partial^2_\theta l_W(\theta | \mathbf{x}_W)\right|_{\theta={\theta}_\text{est}}.
\end{align}
Fig.~\ref{sfig:sample} shows that, in a large sample size, the estimation of error closes to the theoretical predictions. In this Figure, we demonstrate the contextuality-enabled enhancement for $\theta$ and $\phi$ estimation. We apply a systematic error model to make the theoretical predictions. Explanation for the error model is presented in the next section.

\section{Error and noise analysis}

\subsection{Systematic error model}
We characterize systematic errors of our optical setting. The errors do not make our quantum system to lose its coherence, but they cause unwanted drift in experimental results. However, it is impossible to avoid misalignment of control parameters in experiment, so that one performs experiment by allowing a small amount of systematic errors. For example, the axes of wave-plates can be deviated ~$0.1$ degrees in their alignment.

To anlyze the effect of systematic error, we model a quantum probe state $\hat{\varrho}$ with Bloch representation as
\begin{equation}
    \hat{\varrho}(\vec{r}_s)=\frac{1}{2}\left(I + \vec{r}_s\cdot\vec{\sigma}\right),
\end{equation}
where $\vec{r}_s=r_s(\sin\theta_s \cos\phi_s,\sin\theta_s \sin\phi_s,\cos\theta_s)$ satisfying $\abs{r_s}\le1$ and $\vec{\sigma}=(\sigma_x,\sigma_y,\sigma_z)$. For an element of POVM $\hat{m}$ of a binary outcome $a$, we use the representation including biasedness $x$ and unsharpness $\mu$ of outcome, which is given by
\begin{equation}
    \hat{m}(a|x,\vec{\mu})=\frac{1}{2}\left\{\left[1+(-1)^a x\right]I + (-1)^a\vec{\mu}\cdot\vec{\sigma}\right\},
\end{equation}
where $\vec{\mu} = \mu(\sin\theta \cos\phi,\sin\theta \sin\phi,\cos\theta)$ for $0\le\mu\le1$, and the condition for the measurement to be a POVM is $\abs{x} \le 1-\mu$~\cite{Yu2010}. We also represent overall drift of measurement parameters with a linear model
\begin{eqnarray}
    \theta = \theta_0 + \theta_1 \theta_\text{exp},\quad \text{and} \quad
    \phi = \phi_0 + \phi_1 \phi_\text{exp},
\end{eqnarray}
where $\theta_0$ and $\phi_0$ represent shift of parameters, and $\theta_1$ and $\phi_1$ represent scaling of parameters. Finally, a probability obtained by the probe state and the measurement model is given by
\begin{eqnarray}
    p(a|t_s,t) &=& \text{tr}\hat{\varrho}(\vec{r}_s)\hat{m}(a|x,\vec{\mu}) \nonumber \\
          &=& \frac{1}{2}\left\{ \left[1+(-1)^a x\right] + \vec{r}_s\cdot\vec{\mu} \right\},
\end{eqnarray}
where we abbreviate the model parameters as $t_s=\vec{r}_s$ and $t=(x,\vec{\mu})$.

Our goal is to find the model parameters that best describe the experiments. We collect empirical frequencies for each outcome $a$ for a set of states $R =\{\vec{r}^i_s|i=1,2,\cdots,N_s\}$ with a priori probability $p(\vec{r}^i_s)$. By Bayes' theorem, the joint probability $p(b,t_s|t)$ is given by
\begin{equation}
    p(b,t_s|t) = p(b|t^i_s,t)p(t^i_s).
\end{equation}
A uniform distribution for the parameters of quantum state is defined by $p(t^i_s)=1/4\pi$ as $\int dt_s 1/4\pi = 1$. In terms of $\theta_s$ and $\phi_s$, the uniform distribution comes to another distribution $\sin\theta_s/4\pi$. This distribution on continuous space $0\le\theta_s\le\pi$ and $0\le\phi_s \le2\pi$ approximates to the distribution on discrete lattices of $\Delta\theta_s\Delta\phi_s$ as
\begin{equation}
    p(\theta_s,\phi_s) \approx \Delta\theta_s\Delta\phi_s p(\theta^i_s,\phi^j_s),
\end{equation}
where $(i,j)$ is the index pair to the lattices and $p(\theta^i_s,\phi^j_s)=\sin\theta^i_s/N$. Here, $N_s$ is a normalization constant,
\begin{equation}
    N_s = \sum_{i,j}\sin\theta^i_s=N_{\phi_s}N_{\theta_s}\langle\sin\theta_s\rangle = \frac{2N_{\phi_s}N_{\theta_s}}{\pi}\left(\frac{1}{2}\sum_i\frac{\pi}{N_{\theta_s}}\sin\theta^i_s \right) ~\rightarrow~ \frac{2N_{\phi_s}N_{\theta_s}}{\pi}~\text{as}~N_s\rightarrow\infty
\end{equation}
for the lattice of $N_{\phi_s} \times N_{\theta_s}$. We use $\lim_{N_{\theta_s}\rightarrow\infty}\sum_i{\pi}\sin\theta^i_s / {N_{\theta_s}}= \int^\pi_0 d\theta_s\sin\theta_s = 2$. With the a priori distribution on the discrete lattice, our model for the joint probability becomes
\begin{eqnarray}
    p(a,\theta^i_s,\phi^j_s|t) = p(a|\theta^i_s,\phi^j_s,t)p(\theta^i_s,\phi^j_s) = \frac{\sin\theta^i_s}{2N_s}\left\{ \left[1+(-1)^a x\right] +(-1)^a \vec{r}_s^{ij}\cdot\vec{\mu} \right\}.
\end{eqnarray}

To estimate the parameter $t$ of measurement model that best describes the experimental results $N(a|t^{ij}_s)$ for each lattice point $(\theta^i_s,\phi^j_s)$, we use Bayesian inference. We define a log-likelihood function as
\begin{eqnarray}
    l(t)&:=& \sum_{a,i,j}f(a|t^{ij}_s)p(t^{ij}_s)\log p(a|t^{ij}_s,t)p(t^{ij}_s) \nonumber \\
    &=&\sum_{a,i,j}f(a,t^{ij}_s)\log p(a,t^{ij}_s|t)
\end{eqnarray}
where $f(a|t^{ij}_s)=N(a|t^{ij}_s)/N(t^{ij}_s)$ and $N(t^{ij}_s)=\sum_a N(a|t^{ij}_s)$. An estimate is a value to give an extremum of log-likelihood function $l(t)$. We assume that the probabilities given by the model is a baseline. In this case, finding an extreme value of likelihood function becomes maximizing Kullback–Leibler divergence $D_\text{KL}$, which is defined by
\begin{eqnarray}
    D_\text{KL}\left[f(A|t_s)p(t_s)\Vert p(A|t_s,t)p(t_s)\right]
    &:=& \sum_{a,i,j} f(a,t^{ij}_s) \log \frac{f(a,t^{ij}_s)}{p(a,t^{ij}_s|t)}.
\end{eqnarray}
The $D_\text{KL}$ is zero if the experimental frequency $f(A,t_s)$ is equivalent to the model probability $p(a,t_s|t)$. The two approaches result in the same extreme value as
\begin{eqnarray}
    0 &=& \partial_t D_\text{KL}\left[f(A|t_s)p(t_s)\Vert p(A|t_s,t)p(t_s)\right] \nonumber\\
    &=& \partial_t \sum_{a,i,j} f(a|t^{ij}_s)p(t^{ij}_s) \log \frac{f(a|t^{ij}_s)p(t^{ij}_s)}{p(a|t^{ij}_s,t)p(t^{ij}_s)} \nonumber\\
    &=& -\partial_t l(t).
\end{eqnarray}

As the coQM utilizes the two measurements $A$ and $B$, we collect empirical frequencies for each measurement, and consider a minimization problem to analyze the systematic errors:
\begin{eqnarray}
    &&\underset{t_A,t_B}{\text{minimize}}~~~
    D_{KL}\left[f(A,t_s)\Vert p(A,t_s|t_A)\right]+
    D_{KL}\left[f(B,t_s)\Vert p(B,t_s|t_B)\right]\nonumber\\
    &&\text{subject to}~~ \abs{x_A}\le 1-\mu_A,~ 0\le \mu_A\le1,~\abs{x_B}\le 1-\mu_B,~\text{and}~0\le\mu_B\le1.
\end{eqnarray}
In summary, we characterize the measurements $A$ and $B$ with the model parameters $t_A=(x_A,\mu_A,\theta_0,\theta_1,\phi_0,\phi_1)$ and $t_B=(x_B,\mu_B,\theta_0,\theta_1,\phi_0,\phi_1)$, where meaning of each parameter is following;\par
\makebox[1.5cm]{$\theta_0$:}overall drift of $\theta$ \par
\makebox[1.5cm]{$\phi_0$:}overall drift of $\phi$ \par
\makebox[1.5cm]{$\theta_1$:}scaling factor of $\theta$ \par
\makebox[1.5cm]{$\phi_1$:}scaling factor of $\phi$\par
\makebox[1.5cm]{$x_{A,B}$:}biasedness of measurement $A$, $B$\par
\makebox[1.5cm]{$\mu_{A,B}$:}sharpness of measurement $A$, $B$\par
\makebox[1.5cm]{$\theta_{A,B}$:}shift of $\theta$ for measurement $A$, $B$\par
\makebox[1.5cm]{$\phi_{A,B}$:}shift of $\phi$ for measurement $A$, $B$\par

The parameter values for the ideal (no error) case and for an exemplary case with sample size $N_s=10^5$ are shown in the table~\ref{stab:table}. We apply these model parameters to make the theoretical prediction in Fig.~\ref{sfig:sample}.

\begin{table*}[h!]
\caption{\label{stab:table} The model parameters to describe the systematic errors.}
\begin{ruledtabular}
\begin{tabular}{ccccccccccccc}
&$\theta_0$ &$\phi_0$ &$\theta_1$ &$\phi_1$&$x_A$&$\mu_A$&$\theta_A$&$\phi_A$&$x_B$&$\mu_B$&$\theta_B$&$\phi_B$\\
\hline
No error&$0.0$&$0.0$ &$1.0$&$1.0$&$0.0$&$1.0$&$0.0$&$0.0$&$0.0$&$1.0$&$0.5$&$0.0$ \\
Experiment&$0.00023$&$0.0078$&$1.0$&$1.0$&$-0.0015$&$1.0$&$0.0$&$0.0$&$0.0016$&$0.99$&$0.5$&$0.0$\\
\end{tabular}
\end{ruledtabular}
\end{table*}

\begin{figure*}[t!]
    \centering
    \includegraphics[width=\textwidth]{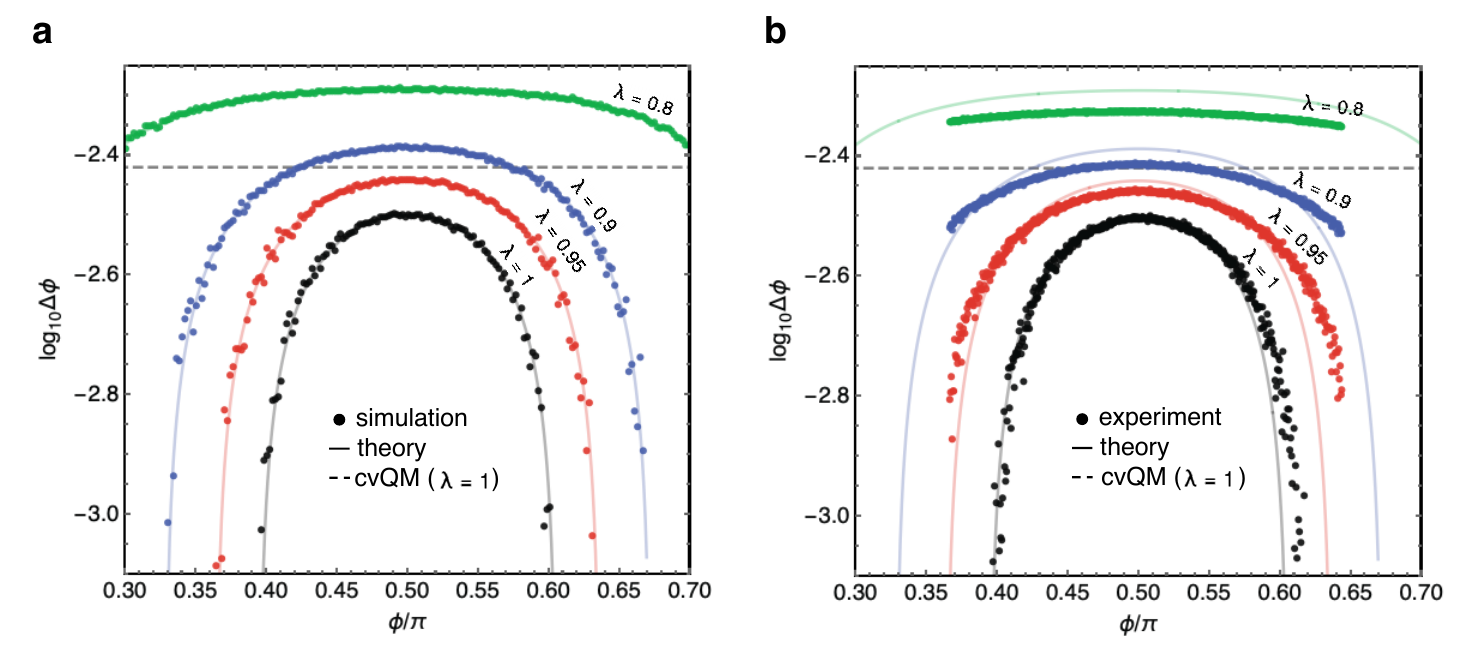}
    \caption{{\bf Effect of depolarization noise on contextual quantum metrology for $\phi$ estimation} Estimation error of contextual quantum metrology changes depending on the purity of initial probe state $\lambda$. We prepare the initial states as $\ket{\psi}=\cos\theta/2\ket{H} + e^{i\phi}\sin\theta/2$ for $\theta=\pi/5$ and $0.4\pi \le \phi \le 0.6\pi$, and assume that they are pure states. We postprocess the probabilities of measurements $A$ and $B$ by following the depolarization model in Eq.~\eqref{seq:depol}. The {\bf a} and {\bf b} show the results of Monte-Carlo simulation and experiment, respectively. Both cases use $2\times10^5$ samples. The dashed lines are the minimum error in the conventional quantum metrology (cvQM). This bound is given by quantum Fisher information of the pure probe states $\ket{\psi}$s. The solid lines are theoretical predictions to which the systematic error model is not applied.}
    \label{sfig:depol}
\end{figure*}

\subsection{Effect of depolarization noise}
\label{sec:noise}
%Purity of quantum states can be degraded in various manners. 
We investigate how performance of contextual quantum metrology depends on purity of a probe state. Depolarization noise degrades coherence of a quantum system, so it is severe error to the quantum system. We represent a partially depolarized quantum state $\hat{\varrho}_d$ as 
\begin{equation}
%\hat{\varrho} = \lambda\hat{\varrho}_0 + \frac{(1-\lambda)}{3}I,
\hat{\varrho}_d = \lambda\hat{\varrho}_0 + \frac{(1-\lambda)}{3}\left({\sigma}_{x}\hat{\varrho}_0{\sigma}_{x} + {\sigma}_{y}\hat{\varrho}_0{\sigma}_{y} + {\sigma}_{z}\hat{\varrho}_0{\sigma}_{z}\right),
\end{equation}
where $\hat{\varrho}_0$ is a pure probe state, ${\sigma}$s are Pauli matrices, and $\lambda\in[0.25,1]$ indicates degree of purity. $\lambda = 1$ for a pure state and $\lambda = 0.25$ for a fully depolarized state $I/2$, where $I$ is an identity matrix. To model the outcome probabilities of partially depolarized state $\hat{\varrho}_d$, we mix probabilities obtained by measuring the states $\hat{\varrho}_0$ and $\hat{\varrho}_{i\in\{x,y,z\}} = \sigma_i\hat{\varrho}_0\sigma_i$ with a measurement as
\begin{equation}
\label{seq:depol}
    P_d(a,b) = \lambda P_0(a,b) + \frac{(1-\lambda)}{3}\left[P_x(a,b) + P_y(a,b) + P_z(a,b)\right],
\end{equation}
where $P_i$ is the probability obtained by measuring the state $\hat{\varrho}_i$.

For $\phi$ estimation, we investigate how the estimation error changes by varying $\lambda$ from $1$ to $0.8$. Monte-Carlo simulation (Fig.~\ref{sfig:depol} {\bf a}) shows the degradation of the estimation precision by the depolarization. The gray dashed line is the minimum error in the conventional quantum metrology (cvQM) obtained by the pure probe state. In experiment (Fig.~\ref{sfig:depol} {\bf b}), we assume that our probe states are pure. We observe the contextuality-enabled enhancement if $\lambda=1$, $0.95$ and $0.9$. If $\lambda=0.8$, the enhancement is disappeared in the range of the parameters we consider.

\end{document}